\documentclass[a4paper]{article}
\usepackage[a4paper,top=3cm,bottom=2cm,left=3cm,right=3cm,marginparwidth=1.75cm]{geometry}
\usepackage{amsfonts}
\usepackage{bm}
\usepackage{extarrows}
\usepackage{amssymb}
\usepackage{color}
\usepackage[all]{xy}
\usepackage{graphicx}
\usepackage{braket}
\usepackage{amsmath}
\usepackage{appendix}
\usepackage{graphicx}
\usepackage[colorinlistoftodos]{todonotes}
\usepackage{amsthm}
\usepackage{mathrsfs}
\usepackage{amssymb}
\usepackage{accents}
\usepackage{extarrows}
\usepackage{makecell}
\usepackage{authblk}
\usepackage{url}
\usepackage{hyperref}
\usepackage{cite}
\usepackage{eufrak}
\hypersetup{colorlinks,
            citecolor=black,
            linkcolor=black,
            urlcolor=green,
            pdftex}

\usepackage[english]{babel}
\usepackage[utf8x]{inputenc}
\usepackage[T1]{fontenc}

\usepackage[normalem]{ulem}

\title{Parametrization of holonomy-flux phase space in the Hamiltonian formulation of $SO(N)$ gauge field theory with $SO(D+1)$ loop quantum gravity as an exemplification}
\author[1,2]{Gaoping Long \footnote{201731140005@mail.bnu.edu.cn}\thanks{corresponding author}}

\affil[1]{Department of Physics, Jinan University, Guangzhou 510632, China}
\affil[2]{Department of Physics, Beijing Normal University, Beijing 100875, China}

\date{}

\begin{document}

\maketitle

\begin{abstract}
The $SO(N)$ Yang-Mills gauge theory is concerned since it can be used to explore the new theory beyond the standard model of particle physics and the higher dimensional loop quantum gravity. The canonical formulation and loop quantization of $SO(N)$ Yang-Mills theory suggest a discrete $SO(N)$ holonomy-flux phase space, and the properties of the critical quantum algebras in the loop quantized $SO(N)$ Yang-Mills theory are encoded in the symplectic structure of this $SO(N)$ holonomy-flux phase space. With
the $SO(D+1)$ loop quantum gravity as an exemplification of loop quantized $SO(N)$ Yang-Mills gauge theory, we introduce
 a new parametrization of the $SO(D+1)$ holonomy-flux phase space in this paper. Moreover, the symplectic structure of the $SO(D+1)$ holonomy-flux phase space are analyzed in terms of the parametrization variables. Comparing to the Poisson algebras among the $SO(D+1)$ holonomy-flux variables, it is shown that the Poisson algebras among the parametrization variables take a clearer formulation, i.e., the Lie algebras of $so(D+1)$ and the Poisson algebras between angle-length pairs.

\end{abstract}

\section{Introduction}
Many of powerful and fundamental theories in modern physics are formulated as the Yang-Mills gauge field theories \cite{PhysRev.96.191}, e.g. Quantum electroweak theory is based on the $SU(2)_{\text{L}}\times U(1)_{\text{Y}}$ gauge theory \cite{GLASHOW1961579,SALAM1964168,PhysRevLett.19.1264}, and Quantum chromodynamics (QCD) takes the formulation of $SU(3)$ gauge theory \cite{PhysRevLett.31.494,PhysRevLett.30.1343}. These theories provide the basement of the Standard Model (SM) of particle physics.  Besides, as a non-perturbative and background-independent approach to unify general relativity (GR) and quantum
mechanics, loop quantum gravity (LQG) in $(1+3)$-dimensional spacetime is developed based on $SU(2)$ gauge theory \cite{Ashtekar2012Background}\cite{Han2005FUNDAMENTAL}\cite{thiemann2007modern}\cite{rovelli2007quantum}, and it has made remarkable progresses in several aspects. For instance, various symmetry-reduced models are established in the framework of LQG to give the resolution of singularities \cite{Ashtekar_2011,Zhang:2021zfp,Ashtekar:2018cay}, and various attempts are made in the framework of
LQG to account for the BH entropy \cite{Engle:2009vc,Agullo:2008yv,Song:2020arr,Song:2022zit}. Nevertheless, several shortcomings of the SM of particle physics suggest new theories beyond the SM, and one of strategy to this goal is considering the gauge field theory with the higher rank gauge group, i.e. the grand unified theories with gauge group $SU(5)$ or $SO(10)$ \cite{PhysRevLett.32.438,FRITZSCH1975193}. Moreover, the loop quantum gravity is extended to higher dimensional spacetime to explore its potential to unify other interactions (e.g.,  extra dimensions and the Kaluza-Klein idea in loop quantization framework \cite{KK2001,Overduin:1997sri}), and $SO(D+1)$ gauge theory is considered in the general $(1+D)$-dimensional LQG \cite{bodendorfer2013newi}.
All of these utilization of $SO(N)$ gauge theory suggest that it is worth to study the explicit structure of the gauge group $SO(N)$ and its Lie algebra. Specifically, in the loop quantization framework of $SO(N)$ gauge theory, the information of the critical quantum algebras is encoded in the symplectic structure of the discrete $SO(N)$ holonomy-flux phase space \cite{Ashtekar2012Background,Han2005FUNDAMENTAL,bodendorfer2013newiii,PhysRevD.11.395}. Nevertheless, this symplectic structure for general $SO(N)$ is rather complicated so that it is difficult to separate out the degrees of freedom that is needed. In this article, a parametrization of the $SO(N)$ holonomy-flux phase space will be introduced, and we will analyze the symplectic structure on and the Poisson structure in terms of the parametrization variables. More explicitly, we will consider  $(1+D)$-dimensional LQG with gauge group $SO(D+1)$ as an example of $SO(N)$ gauge theory,  and then introduce the parametrization of the $SO(D+1)$ holonomy-flux phase space. In the following part of this article, this parametrization will be referred as to twisted geometry parametrization, since it has a discrete geometry interpretation by imposing some constraints in LQG framework.

Now, let us turn to the $(1+D)$-dimensional LQG which is a representative example of $SO(N)$ gauge theory. The loop quantization approach for GR in all dimensions is first developed by Bodendorfer, Thiemann and Thurn \cite{bodendorfer2013newi}\cite{bodendorfer2013newiii}\cite{bodendorfer2013implementation}.  In detail, the all dimensional LQG  is based on the connection formulation of $(1+D)$ dimensional GR in the form of the $SO(D+1)$ Yang-Mills theory, with the kinematic phase space coordinatized by the canonical pairs $(A_{aIJ},\pi^{bKL})$, consisting of the spatial $SO(D+1)$ connection fields $A_{aIJ}$ and the vector fields $\pi^{bKL}$. In this formulation, the theory is governed by the first class system of the $SO(D+1)$ Gaussian constraints, the $(D+1)$-dimensional ADM constraints and the additional simplicity constraints. Similar to the Gaussian constraints, the simplicity constraints taking the form $S^{ab}_{IJKL}:=\pi^{a[IJ}\pi^{|b|KL]}$ generate extra gauge symmetries in the $SO(D+1)$ Yang-Mills phase space. It has been shown that the connection phase space correctly reduces to the familiar ADM phase space by carrying out the symplectic reductions with respected to the Gaussian and simplicity constraints. Similar to the case of the $SU(2)$ LQG, the loop quantization of the $SO(D+1)$ Yang-Mills theory leads to the spin-network states of the $SO(D+1)$ holonomies on some graphes, which carry the quanta of the flux operators representing the fluxes of $\pi^{bKL}$ over some $(D-1)$-dimensional faces. The Hilbert space composed by the spin-network states indicates the holonomy-flux phase space associated to each graph, with the Poisson algebras among holonomies and fluxes in the holonomy-flux phase space being isomorphic to the quantum algebras among them in the quantum Hilbert space. To look for the all-dimensional Regge ADM data encoded in the $SO(D+1)$ spin-network states, it is necessary to find the degrees of freedom of discrete geometries encoded in the $SO(D+1)$ holonomy-flux variables, by considering a gauge reduction procedure with respect to both of the $SO(D+1)$ Gaussian constraints and the simplicity constraints in the holonomy-flux phase space.

A series of studies in this direction is first carried out in the $SU(2)$ formulation of $(1+3)$-dimensional LQG \cite{Freidel_2010}\cite{bianchi2011polyhedra}\cite{Conrady_2009}\cite{freidel2010twisted}\cite{Bianchi:2009ky}, and then they are generalized to the $SO(D+1)$ holonomy-flux phase space in our companion paper \cite{Long:2020tlr,Long:2020agv}. Specifically, since the simplicity constraints become anomalous at the vertices of the graphs, the reductions with respect to the  Gaussian and simplicity constraints are guided by the twisted geometry parametrization of the edge simplicity constraint surface in the holonomy-flux phase space of $SO(D+1)$ LQG.
 Especially, the twisted geometry interpretation of holonomy-flux variables suggests that the Gaussian and edge simplicity constraints should be imposed strongly since they generate true gauge transformations, while the vertex simplicity constraints should be imposed weakly. The reduced space parametrized by the twisted geometric parameters give a discrete Regge geometry picture, which can be regarded as the discrete version of the ADM phase space of GR.
An important application of the twisted geometry parametrization is the construction of the twisted geometry coherent state. Such kind of coherent states is firstly established in $SU(2)$ LQG \cite{Calcinari:2020bft}, and then it is generalized to the $SO(D+1)$ LQG with the restriction of the simple representations \cite{Gaoping2019coherentintertwiner,Long:2020euh,Long:2021xjm,Long:2021lmd,Long:2022cex}. Specifically, based on the twisted geometry parameters, the simple twisted geometry coherent state in the strong solution space of quantum edge simplicity constraints is established by selecting the dominant terms (which is referred to as Perelomov type coherent state \cite{Long:2020euh,perelomov2012generalized,Long:2020agv}) with simple representation of $SO(D+1)$ in the decomposition of the heat-kernel coherent state of $SO(D+1)$ \cite{Thiemann_2001,Thiemann_20012,Thiemann_20013}. It has been shown that the simple twisted geometry coherent states take the Gaussian superposition formulations. Especially, the simple twisted geometry coherent states provides an over-complete basis of the strong solution space of quantum edge simplicity constraints, and their wave functions have well-behaved peakedness and Ehrenfest properties in the reduced phase space with respect to the edge simplicity constraints \cite{Long:2021lmd,Long:2022cex}.

In fact, the twisted geometry  parametrization of the $SO(D+1)$ holonomy-flux phase space discussed in Ref.\cite{Long:2020tlr} concerns the issues on the constraint surface of edge simplicity constraint, and the resulted twisted geometry variables only give the parametrization of the reduce phase space with respect to edge simplicity constraint. Correspondingly, the simple twisted geometry coherent states constructed based on the twisted geometry parametrization of the reduce phase space are the gauge (with respect to edge simplicity constraint) invariant coherent states \cite{Long:2021lmd}. In other words, the wave functions of these gauge (with respect to edge simplicity constraint) invariant coherent states are constants along the corresponding gauge orbits, so that each of them peaks at a gauge orbit instead of a point in the phase space \cite{Long:2022thb}.

As we have mentioned above, the edge simplicity constraint should be imposed strongly following the twisted geometry interpretation of holonomy-flux variables. Thus, it seems that all of the studies for all dimensional $SO(D+1)$ LQG can be completed in the strong solution space of quantum edge simplicity constraint, which is the gauge (with respect to simplicity constraint) invariant subspace of the full Hilbert space of all dimensional $SO(D+1)$ LQG.  Nevertheless, several discussions has shown that it is necessary to consider some kind of gauge fixed solution space with respect to simplicity constraint, to deal with some of the issues appeared in the  all dimensional $SO(D+1)$ LQG.
Let us introduce two issues to explain this necessity. First, the regularization of the scalar constraint can be carried out by following the standard loop regularization method \cite{bodendorfer2013newi}\cite{bodendorfer2013newiii}\cite{Gaoping2019geometricoperators}. The resulted regularized scalar constraint contains the Euclidean term which is given by the antisymmetric contraction of the holonomies along some closed loops and the fluxes at the beginning and target point of these loops. Classically, this Euclidean term captures the information of both of the intrinsic and extrinsic curvature along these closed loops. However, it is shown that the Euclidean term in the quantized scalar constraint can not capture the information of those intrinsic and extrinsic curvature in the strong solution space of quantum edge simplicity constraint, since the strong imposition of quantum edge simplicity constraint leads to the gauge averaging, which vanishes some critical ingredients in the holonomies \cite{Long:2022thb}. Thus, the standard loop regularization method is conflict to the strong imposition of the edge simplicity constraint. To deal with issue, one may consider the gauge fixed solution of the edge simplicity constraint to avoid the gauge averaging, so that the scalar constraint operator given by standard loop regularization method captures the information of those intrinsic and extrinsic curvature correctly. This is the first issue which points out the necessity to consider then gauge fixed solution space with respect to simplicity constraint. The second issue which points out this necessity is the the Fermion coupling problem in all dimensional LQG \cite{Bodendorfer:2011ny,Bodendorfer:2011pb}. Specifically, the strong imposition of the quantum edge simplicity constraint restricts that the holonomies in all dimensional LQG can only be represented in the simple representation space of $SO(D+1)$, which leads that the holonomies can not transform the Fermions which take values in the spinor representation space of $SO(D+1)$ for $D\geq4$. An alternative scheme to deal with this issue is to consider the gauge fixed solution of quantum edge simplicity constraint based on the coherent states, which ensures that the holonomies could take matrixes in the spinor representation space of $SO(D+1)$, so that they are able to describe the transformation of Fermions along edges.

 Usually, in the classical theory, the gauge fixing can be realized by restricting the physical considerations on  a section of the gauge orbits on the constraint surface of edge simplicity constraint. However, this is not valid in the quantum theory, since the wave functions of the quantum states which sharply converge to the constraint surface of edge simplicity constraint are always dispersed along the gauge orbits.
To overcome this problem, it is  reasonable to consider the coherent state whose wave function peaks at a point in the phase space, so that one could have the state whose wave function converges to both of the constraint surface of edge simplicity constraint and a section of the gauge orbits, with this convergence is controlled by the width of the wave function of the coherent state.
Such kind of coherent state whose wave function peaks at a point in the $SO(D+1)$ holonomy-flux phase space could be constructed by following a similar procedure as the construction of the simple twisted geometry coherent state in the strong solution space of quantum edge simplicity constraint \cite{Long:2021lmd}. More specifically, one need to consider a more generalized twisted geometry parametrization, which is able to coordinate the (almost) whole $SO(D+1)$ holonomy-flux phase space instead of the reduced phase space. Then, based on this more generalized twisted geometry parametrization, one could decompose the heat-kernel coherent state of $SO(D+1)$ and select some dominant terms to formulate the twisted geometry coherent state involving the non-simple representations of $SO(D+1)$, which will be referred as to the non-simple twisted geometry coherent state in all dimensional LQG.

As the preparation step to establish the non-simple twisted geometry coherent state in all dimensional LQG, it is necessary to extend the twisted geometry parametrization to the full $SO(D+1)$ holonomy-flux phase space. In this article, we will establish the twisted geometry parametrization of the full $SO(D+1)$ holonomy-flux phase space, and extend this parametrization as a symplectic-morphism. Besides, we will show that  the twisted geometry parametrization of
edge simplicity constraint surface introduced in our previous work \cite{Long:2020tlr} can be regarded as a special cases of the construction in this article.

This article is organized as follows.  In our brief review of the classical connection formulation of all dimensional GR in Section \ref{sec2}, we will also introduce the $SO(D+1)$ holonomy-flux phase space and the discretized formulation of the kinematical constraints. In Section \ref{sec3} and Section \ref{sec4} we will introduce the twisted geometry parametrization for the full $SO(D+1)$ phase space, and analyze the Poisson structures among the new geometric parametrization variables. Then, in Section \ref{sec5} we will discuss the relation between the twisted geometry parametrizations of the edge simplicity constraint surface and the full $SO(D+1)$ holonomy-flux phase space. Finally, we will conclude with the outlook for the possible next steps of the future research.

\section{Phase space of $SO(D+1)$ loop quantum gravity}\label{sec2}
\subsection{Connection phase space}

The classical connection formulation of GR with arbitrary spacetime dimensionality of $(1+D)$ is first developed by Bodendofer, Thiemann and Thurn in Ref.\cite{bodendorfer2013newi}. This continuum connection phase space is coordinatized by a $so(D+1)$ valued 1-form field $A_{aIJ}$ and a vector field $\pi^{bKL}$ on the $D$-dimensional spatial manifold $\Sigma$, with the non-trivial Poisson brackets between them being given by
\begin{equation}\label{Poisson1}
\{A_{aIJ}(x), \pi^{bKL}(y)\}=2\kappa\beta\delta_a^b\delta_{[I}^K\delta_{J]}^L\delta^{(D)}(x-y),
\end{equation}
where $\beta$ is the Barbero-Immirzi parameter and $\kappa$ is the gravitational constant. It is known that this connection phase space correctly reduces to the familiar ADM phase space after the standard symplectic reduction procedure with respect to the first-class constraint system composed by the Gauss
constraints $\mathcal{G}^{IJ}\approx0$ and simplicity constraints $S^{ab}_{IJKL}:=\pi^{a[IJ}\pi^{|b|KL]}\approx0$. Specifically, the simplicity constraint can be solved as $\pi^{aIJ}=2\sqrt{q}{n}^{[I}e^{|a|J]}$, where $e^a_I$ is a dual D-bein field, ${n}^I$ satisfying ${n}^I{n}_I=1$ is determined by $e^a_I$ with ${n}^Ie_{aI}=0$, and $q$ is the determinant of the spatial metric $q_{ab}$  which is determined by $\pi^{aIJ}$ with $q^{ab}=e^{aI}e^b_I$ on the simplicity constraint surface. One can split $A_{aIJ}$ as
\begin{equation}\label{splitA}
A_{aIJ}\equiv\Gamma_{aIJ}(\pi)+\beta K_{aIJ}
\end{equation}
 where $\Gamma_{aIJ}(\pi)$ is a functional of $\pi^{aIJ}$ and it satisfies $\Gamma_{aIJ}(\pi)=\Gamma_{aIJ}(e)$ on the simplicity constraint surface, with $\Gamma_{aIJ}(e)$ being the unique torsionless spin connection compatible with the D-bein $e_{aI}$. Then, the densitized extrinsic curvature can be given by $\tilde{K}_a^{\ b}=K_{aIJ}\pi^{bIJ}$ on the constraint surface of both Gaussian and simplicity constraint surface.

It is easy to check that the Gaussian constraint generate the standard $SO(D+1)$ gauge transformation of the connection field and its conjugate momentum. Now, let us consider the simplicity constraints from the perspectives of the corresponding gauge transformations. First, the solutions $\pi^{aIJ}=2\sqrt{q}{n}^{[I}e^{|a|J]}$ to the simplicity constraint introduced above defines the constraint surface of the simplicity constraints. Then, one can verify that the infinitesimal gauge transformations induced by simplicity constraints are given by \cite{bodendorfer2013newi}
\begin{equation}\label{gaugetranssim}
\delta K_{c}^{PQ}=\{\int_{\Sigma}d^Dxf_{ab}^{IJKL}\pi^a_{[IJ}\pi^b_{KL]}(x), K_{c}^{PQ}(y)\}=4\kappa f_{cb}^{[PQKL]}\pi^b_{KL}(y).
\end{equation}
Notice that, on the simplicity constraint surface we have $\pi^{aIJ}=2\sqrt{q}{n}^{[I}e^{|a|J]}$ so that
$\delta K_{c}^{IJ}{n}_I=0$. Further, by introducing the decomposition
\begin{equation}
K_{aIJ}\equiv 2{n}_{[I}K_{|a|J]}+\bar{K}_{aIJ},
\end{equation}
where $\bar{K}_{aIJ}:=\bar{\eta}_I^K\bar{\eta}_J^LK_{aKL}$ with $\bar{\eta}^I_J=\delta^I_J-{n}^I {n}_J$ and $\bar{K}_{aIJ}{n}^I=0$, we immediately find that $\bar{K}_{aIJ}$ is the pure gauge component, while the components $2{n}_{[I}K_{|a|J]}$ are gauge invariant with respect to the  transformations given in (\ref{gaugetranssim}). From the expressions of the ADM variables $qq^{ab}=\frac{1}{2}\pi^{aIJ}\pi^b_{IJ}$ and $\tilde{K}_a^{\ b}=K_{aIJ}\pi^{bIJ}$, it is easy to see that these variables are indeed gauge invariant with respect to the simplicity constraints on the constraint surface. Thus, through the symplectic gauge reduction procedure, the simplicity constraints eliminate the two parts of degrees of freedom--- restricting $\bar{\pi}^{aIJ}:=\pi^{aIJ}-2\sqrt{q}n^{[I}e^{|a|J]}=0$ by the constraint equation and removing the pure-gauge components $\bar{K}_{aIJ}:=\bar{\eta}_I^K\bar{\eta}_J^LK_{aKL}$. Following these results, the geometric variables constructed by the ADM variables $(q_{ab},\tilde{K}^{cd})$ can be extended as functionals in the connection phase space, with their original geometric interpretation are remained on the constraints surface.

\subsection{Holonomy-flux phase space}

The quantization of the connection formulation of $(1+D)$-dimensional GR can be carried out by following the standard loop quantization procedures, which leads to a Hilbert space $\mathcal{H}$ given by the completion of the space of cylindrical functions on the quantum configuration space \cite{bodendorfer2013newiii}. This Hilbert space $\mathcal{H}$ can be regarded as a union of the spaces $\mathcal{H}_\gamma=L^2((SO(D+1))^{|E(\gamma)|},d\mu_{\text{Haar}}^{|E(\gamma)|})$ on all possible graphs $\gamma$,  where $E(\gamma)$ denotes the set of edges of $\gamma$ and $d\mu_{\text{Haar}}^{|E(\gamma)|}$ denotes the product of the Haar measure on $SO(D+1)$. The Gaussian constraint and simplicity constraint can be promoted as constraint operators in this Hilbert space. However, it has been turned out that the quantum brackets among these constraints give an open and anomalous quantum algebra, which is distinguished with the corresponding  constraint algebra of first class in connection phase space \cite{bodendorfer2013implementation}. Hence, it is necessary to propose a proper treatment of  these quantum constraints, to reduce the gauge degrees of freedom and remain the physical degrees of freedom correctly. A reasonable method to reach this goal is to construct the gauge reductions with respect to Gaussian and simplicity constraints in the holonomy-flux phase space. More specifically,  since the classical constraint algebras in the holonomy-flux phase space are isomorphic to the quantum constraint algebras in the quantum theory, one can treat the Gaussian and simplicity constraints in the holonomy-flux phase space and quantum theory on the same footing. Then, the degrees of freedom reduced in the procedures of the imposition of quantum constraint operators can be reflected in the procedures of the gauge reductions with respect to Gaussian and simplicity constraints in the holonomy-flux phase space. Through this gauge reductions, one can clarify the gauge degrees of freedom and verify that if the treatment of these constraints remains correct physical degrees of freedom. Now, let us first give a brief review of the holonomy-flux phase space.

The quantum geometry of loop quantum gravity is described based on the spatially smeared variables --- the $D$-bein fluxes over $(D-1)-$dimensional faces and connection holonomies over paths--- for the conjugate pairs of elementary variables.  We will focus on the holonomies and fluxes based on one specific graph for the following. The edges of the given graph naturally provide the set of paths for a fixed set of holonomies, and the cell decomposition dual to the graph provides the set of $(D-1)$-faces specifying a fixed set of fluxes. In this setting, the holonomy over one of the edges is naturally conjugating to the flux over the $(D-1)$-face traversed by the edge, with this pair satisfies the smeared version of the Poisson algebra (\ref{Poisson1}),  and thus form a new phase space. More precisely, given the graph $\gamma$ embedded in the spatial manifold, we consider a new algebra given by  the holonomy-flux variables $(h_e, X_e)\in SO(D+1)\times so(D+1)$ over all edges $e$ of $\gamma$. These pairs of variables represent the discretized version of the connection $A_{aIJ}$ and its conjugate momentum $\pi^{bKL}$. Specifically, the holonomy of $A_{aIJ}$ along an edge $e\in\gamma$ defined by
 \begin{equation}
h_e[A]:=\mathcal{P}\exp(\int_eA)=1+\sum_{n=1}^{\infty}\int_{0}^1dt_n\int_0^{t_n}dt_{n-1}...\int_0^{t_2} dt_1A(t_1)...A(t_n),
 \end{equation}
 where $A(t):=\frac{1}{2}\dot{e}^aA_{aIJ}\tau^{IJ}$, $\dot{e}^a$ is the tangent vector field of $e$, $\tau^{IJ}$ is a basis of $so(D+1)$ given by $(\tau^{IJ})^{\text{def.}}_{KL}=2\delta^{[I}_{K}\delta^{J]}_{L}$ in definition representation space of $SO(D+1)$, and $\mathcal{P}$ denotes the path-ordered product.
The flux $X^{IJ}_e$ of $\pi^{aIJ}$ through the $(D-1)$-dimensional face dual to edge $e$ is defined by
 \begin{equation}
 X^{IJ}_e:=-\frac{1}{4\beta a^{D-1}}\text{tr}\left(\tau^{IJ}\int_{e^\star}\epsilon_{aa_1...a_{D-1}}h(\rho^s_e(\sigma)) \pi^{aKL}(\sigma)\tau_{KL}h(\rho^s_e(\sigma)^{-1})\right),
 \end{equation}
 where $a$ is an arbitrary but fixed constant with the dimension of length, $e^\star$ is the $(D-1)$-face traversed by $e$ in the dual lattice of $\gamma$, $\rho_e^s(\sigma): [0,1]\rightarrow \Sigma$ is a path connecting the source point $s_e\in e$ to $\sigma\in e^\star$ such that $\rho_e^s(\sigma): [0,\frac{1}{2}]\rightarrow e$ and $\rho_e^s(\sigma): [\frac{1}{2}, 1]\rightarrow e^\star$. The Poisson algebra between the holonomy-flux variables can be induced from the Poisson bracket \eqref{Poisson1}  between the connection variables, which reads
  \begin{eqnarray}\label{hXP1}
 &&\{h_e, h_{e'}\}=0,\quad\{h_e, X^{IJ}_{e'}\}=\delta_{e,e'}\frac{\kappa}{a^{D-1}} \frac{d}{d\lambda}(e^{\lambda\tau^{IJ}}h_e)|_{\lambda=0}, \\\nonumber
 && \{X^{IJ}_e, X^{KL}_{e'}\}=\delta_{e,e'}\frac{\kappa}{2a^{D-1}}(-\delta^{IK}X_e^{JL}-\delta^{JL }X^{IK}_e+\delta^{IL}X_e^{JK}+\delta^{JK}X_e^{ IL}).
 \end{eqnarray}
  Notice that $h_e\in SO(D+1)$, $X_e^{IJ}\in so(D+1)$ and $SO(D+1)\times so(D+1)\cong T^\ast SO(D+1)$, the new discrete phase space called the holonomy-flux phase space of $SO(D+1)$ loop quantum gravity on a fixed graph, is a direct product of $SO(D+1)$ cotangent bundles. Finally, the complete phase space of the theory is given by taking the union over the holonomy-flux phase spaces of all possible graphs. Similar to the $SU(2)$ case, the phase space coordinated by the holonomy-flux variables $(h_e, X_e)$  of $SO(D+1)$ loop quantum gravity can be regarded as the discretized version of the continuum phase space.

The (discretized) Gaussian and simplicity constraints in the holonomy-flux phase space are constructed in agreement with the corresponding quantum constraints. With $X_{-e}=-h_e^{-1}X_eh_e\equiv \tilde{X}_e$, the (discretized) Gaussian constraints $G_v^{IJ}\approx0$ for each vertex $v\in \gamma$ of the graph take the form  \cite{bodendorfer2013newiii}
\begin{eqnarray}
\label{gaussconstr}
G_v^{IJ}=\sum_{e|s(e)=v}X_e^{IJ}+\sum_{e|t(e)=v}\tilde{X}_e^{IJ}\approx0,
\end{eqnarray}
where $s(e)$ and $t(e)$  denote the source and target points of the oriented edge $e$ respectively. The (discretized) simplicity constraints consist of the edge simplicity constraints $S^{IJKL}_e\approx0$ and vertex simplicity constraints $S^{IJKL}_{v,e,e'}\approx0$, which take the forms  \cite{bodendorfer2013newiii}
\begin{equation}
\label{simpconstr}
S_e^{IJKL}\equiv X^{[IJ}_e X^{KL]}_e\approx0, \ \forall e\in \gamma,\quad S_{v,e,e'}^{IJKL}\equiv X^{[IJ}_e X^{KL]}_{e'}\approx0,\ \forall e,e'\in \gamma, s(e)=s(e')=v.
\end{equation}
It has been shown that, since the commutative Poisson algebra between the conjugate momentum variables $\{\pi^{bKL}\}$ becomes non-commutative Poisson algebra between the flux variables $\{ X^{KL}_e\}$ after the smearing, the Poisson algebra among the discrete version of simplicity constraints become non-closed and thus anomalous, which leads that the symplectic reductions in the holonomy-flux phase space becomes difficult to implement  \cite{bodendorfer2013implementation}. To deal with this issue, the twisted geometry parametrization of the holonomy-flux phase space is constructed, which ensures that the gauge reductions with respect to the Gaussian and simplicity constraint in the holonomy-flux phase space can be carried out with the guidance of the twisted geometric interpretation of the holonomy-flux variables \cite{Long:2020tlr}.

 The twisted geometry parametrization for the the $SU(2)$ holonomy-flux variables of $(1+3)$-dimensional LQG is first introduced by a series of studies following the original works by Freidel and Speziale \cite{Freidel_2010}\cite{freidel2010twisted}. The space of the twisted geometry for $SU(2)$ LQG can undergo a symplectic reduction with respect to the discretized Gauss constraints, giving rise to a reduced phase space containing the discretized ADM data of a polyhedral Regge hypersurface. Following a similar procedure, the twisted geometry parametrization in all dimensional $SO(D+1)$ LQG has been constructed on the edge simplicity constraint surface in the $SO(D+1)$ holonomy-flux phase space in our companion paper \cite{Long:2020tlr}. It has been shown that the gauge reductions with respect to the simplicity constraints and Gaussian constraints in $SO(D+1)$ LQG can be carried out properly in the twisted geometry parametrization space, which leads to a clear correspondence between the original holonomy-flux variables $(h_e, X_e)$ on edge simplicity constraint surface and the $D$-hypersurface discrete geometry data in Regge geometry formulation.  Nevertheless, it is not enough to construct the twisted geometric parametrization  on the edge simplicity constraint surface in the $SO(D+1)$ holonomy-flux phase space.
As we have mentioned in introduction, several explorations in the quantum theory of $SO(D+1)$ LQG requires us consider the quantum states whose wave functions are dispersed beyond the edge simplicity constraint surface. Hence, it is necessary to extend the twisted geometry parametrization to interpret the phase space points which are not located in  the edge simplicity constraint surface.

\section{Geometric parametrization of $SO(D+1)$ holonomy-flux phase space}\label{sec3}
To ensure our statements and the notations clearer, we will first generalize the twisted geometry parametrization to $T^\ast SO(D+1)$ in this section. Then, it will be left to section 5 to discuss the relation between the twisted geometry parametrizations constructed in this article and previous works \cite{Long:2020tlr}.
\subsection{Beyond the edge-simplicity constraint surface}
Recall the $SO(D+1)$ holonomy-flux phase space $\times_{e\in \gamma}T^\ast SO(D+1)_e$ associated to the given graph $\gamma$.  Let us focus on the holonomy-flux phase space $T^\ast SO(D+1)$ associated to a single edge without loss of generality.
To give the explicit formulation of the twisted geometric parametrization of $T^\ast SO(D+1)$, let us first introduce some new notations. Consider the orthonormal basis $\{\delta_1^I,\delta_2^I,...,\delta_{D+1}^I\}$ of $\mathbb{R}^{D+1}$, one has the basis $\{\tau_{IJ}\}$ of $so(D+1)$  given by $\tau_{IJ}=(\tau_{IJ})^{KL}_{\text{def.}}:=2\delta_I^{[K}\delta_J^{L]}$ in the definition representation space of $SO(D+1)$, where $(\tau_{IJ})^{KL}_{\text{def.}}$ is the generator of the infinitely small rotation in the 2-dimensional vector space spanned by the two vectors $\delta_I^K$ and  $\delta_J^L$.
Then, let us introduce the maximum commutative sub-Lie algebra of  $so(D+1)$ spanned by $\{\tau_1, \tau_2,...,\tau_m\}$ with $m=[\frac{D+1}{2}]$, where we define
\begin{equation}
\tau_1:=\tau_{12},\  \tau_2:= \tau_{34},\ ...,  \  \tau_m:= \tau_{D,D+1}
\end{equation}
for $D+1$ being even, and
 \begin{equation}
\tau_1:=\tau_{12},\  \tau_2:= \tau_{34},\ ...,  \  \tau_m:= \tau_{D-1,D}
\end{equation}
for $D+1$ being odd.
This maximum commutative sub-Lie algebra of  $so(D+1)$ generates the maximum commutative subgroup $\mathbb{T}^m$. Then, $SO(D+1)$ can be regarded as a fiber Bundle with the fibers $\mathbb{T}^m$ on the base manifold $\mathbb{Q}_m:=SO(D+1)/\mathbb{T}^m$, which can be also given by $\mathbb{Q}_m=\{\mathbb{V}:=(V_1,...,V_m)|V_\imath=g\tau_{\imath} g^{-1}, \imath\in\{1,...,m\}, g\in SO(D+1)\}$. One can choose a Hopf section $n: \mathbb{Q}_m\mapsto SO(D+1), \mathbb{V}\mapsto n(\mathbb{V})$
 and another Hopf section $\tilde{n}: \tilde{\mathbb{Q}}_m\mapsto SO(D+1), \tilde{\mathbb{V}}\mapsto \tilde{n}(\tilde{\mathbb{V}})$ for the copy $\tilde{\mathbb{Q}}_m$ of $\mathbb{Q}_m$, which satisfy
\begin{equation}
V_1=n(\mathbb{V})\tau_{1}n(\mathbb{V})^{-1},...,V_m=n(\mathbb{V})\tau_{m}n(\mathbb{V})^{-1},
\end{equation}
and
\begin{equation}
\tilde{V}_1=-\tilde{n}(\tilde{\mathbb{V}})\tau_{1}\tilde{n}(\tilde{\mathbb{V}})^{-1}, ...,\tilde{V}_m=-\tilde{n}(\tilde{\mathbb{V}})\tau_{m}\tilde{n}(\tilde{\mathbb{V}})^{-1}
\end{equation}
with $\mathbb{Q}_m\ni\mathbb{V}:=(V_1,...,V_m)$ and $\tilde{\mathbb{Q}}_m\ni\tilde{\mathbb{V}}:=(\tilde{V}_1,...,\tilde{V}_m)$.

Observe that the choice for the Hopf sections is clearly non-unique, and from now on our parametrization will be given under one fixed choice of $\{n_e,\tilde{n}_e\}$ for each edge $e$. Also, we will use the notations $n_e\equiv n_e(\mathbb{V}_e)$ and $\tilde{n}_e\equiv\tilde{n}_e(\tilde{\mathbb{V}}_e)$ in the following part of this article.
Then, in the space $T^\ast SO(D+1)_e$ associated to each edge $e$, the generalized twisted geometry parametrization can be given by the map
\begin{eqnarray}\label{para}
(\mathbb{V}_e,\tilde{\mathbb{V}}_e,\vec{\eta}_e,\vec{\xi}_e)\mapsto(h_e, X_e)\in T^\ast SO(D+1)_e: && X_e=\frac{1}{2}n_e(\eta_e^1 \tau_1+...+\eta_e^m \tau_m)n_e^{-1}\\\nonumber
&&h_e=n_ee^{\xi_e^1\tau_1}...e^{\xi_e^m\tau_m}\tilde{n}_e^{-1},
\end{eqnarray}
where we defined the length parameters $\vec{\eta}_e:=(\eta_e^1,...,\eta_e^m)$, $\eta_e^1,\eta_e^2,...,\eta_e^m\in\mathbb{R}$ with $\eta_e^1\geq\eta_e^2\geq,...,\geq|\eta_e^m|\geq0$ and the angle parameters $\vec{\xi}:=(\xi_e^1,...,\xi_e^m)$ with $\xi_e^1,\xi_e^2,...,\xi_e^m\in(-\pi,\pi]$. Especially,  $\eta_e^m$ satisfies $\eta_e^m\in\mathbb{R}$ for $D+1$ being even, and $\eta_e^m\geq0$ for $D+1$ being odd.
Let us introduce $\chi^1,...,\chi^{m}\geq 0$,  by defining
 \begin{equation}\label{chidef1}
 \eta_e^1=:\chi_e^1+...+\frac{\chi_e^m}{2}, \quad \eta_e^2 =:\chi_e^2+...+\frac{\chi_e^m}{2},
 \end{equation}
 \begin{equation}
\\\nonumber ...,
 \end{equation}
 \begin{equation}
\eta_e^{m-1}=:\chi_e^{m-1}+\frac{\chi_e^m}{2},\quad \eta_e^m=:\frac{\chi_e^m}{2}
 \end{equation}
 for $D+1$ being odd, and
 \begin{equation}
 \eta_e^1=:\chi_e^1+...+\frac{\chi_e^{m-1}}{2}+\frac{\chi_e^m}{2}, \quad \eta_e^2 =:\chi_e^2+...+\frac{\chi_e^{m-1}}{2}+\frac{\chi_e^m}{2},
 \end{equation}
 \begin{equation}
\\\nonumber ...,
 \end{equation}
 \begin{equation}\label{chidef2}
\eta_e^{m-1}=:\frac{\chi_e^{m-1}}{2}+\frac{\chi_e^m}{2},\quad \eta_e^m=:\frac{\chi_e^{m-1}}{2}-\frac{\chi_e^m}{2}
 \end{equation}
for $D+1$ being even,  one  can replacing $\vec{\eta}_e$ by $\vec{\chi}_e:=(\chi_e^1,...,\chi_e^m)$ in the parametrization \eqref{para}.

The twisted geometry parametrization \eqref{para} of  $T^\ast SO(D+1)_e$ associated to a single edge  can be directly extended to the whole graph $\gamma$.
Correspondingly, one can introduce the Levi-Civita holonomies $\{h^{\Gamma}_{e}|e\in\gamma\}$ determined by the fluxes $\{X_e\in so(D+1)|e\in\gamma\}$ and $\{\tilde{X}_e\in so(D+1)|e\in\gamma\}$, which takes the form
\begin{equation}\label{hgamma}
h^{\Gamma}_{e}\equiv n_ee^{\zeta_e^1\tau_1}...e^{\zeta_e^m\tau_m}\tilde{n}_e^{-1}.
\end{equation}
 Note that the variables $(\zeta_e^1,...,\zeta_e^n)$ are well-defined via the given $h^{\Gamma}_{e}$ and the chosen Hopf sections, thus
 $(\zeta_e^1,...,\zeta_e^n)$ are already fixed by the given $\{X_e\in so(D+1)|e\in\gamma\}$ and $\{\tilde{X}_e\in so(D+1)|e\in\gamma\}$. Then, one can factor out $h^{\Gamma}_{e}$ from $h_e$ through the expressions
\begin{eqnarray}
\label{decomp3}
h_e= \left(e^{(\xi_e^1-\zeta_e^1)n_e\tau_1n_e^{-1}}...e^{(\xi_e^m-\zeta_e^m)n_e\tau_mn_e^{-1}}\right)  h^{\Gamma}_{e} =h^{\Gamma}_{e}\left(e^{(\xi_e^1-\zeta_e^1)\tilde{n}_e\tau_1\tilde{n}_e^{-1}}... e^{(\xi_e^m-\zeta_e^m)\tilde{n}_e\tau_m\tilde{n}_e^{-1}}\right)
\end{eqnarray}
in the perspectives of  the source point and target point of $e$ respectively.

 The above decomposition with twisted geometry parameters can be adopted to the splitting of the the Ashtekar connection as $A_a=\Gamma_a+\beta K_a$ on a given graph. Specifically, one can consider the integral of $A_a=\Gamma_a+\beta K_a\in so(D+1)$ along an infinitesimal edge direction $\ell^a_e$, which leads to $A_e\equiv A_a\ell^a_e$, $\Gamma_e\equiv \Gamma_a\ell^a_e$ and $K_e\equiv K_a\ell^a_e$. Clearly, we can establish the following  correspondence of
\begin{eqnarray}
\label{corres1}
h_e= e^{A_e} \,\,\,\text{and}\,\,\,h^{\Gamma}_{e}= e^{\Gamma_e}.
\end{eqnarray}
The remaining factor should account for the $K_e$. According to the above discussion, the value of $K_e$ may thus be expressed in the perspectives of  the source point and target point of $e$, respectively as
\begin{eqnarray}
\label{corres2}
\left(e^{(\xi_e^1-\zeta_e^1)n_e\tau_1n_e^{-1}}...e^{(\xi_e^m-\zeta_e^m)n_e\tau_mn_e^{-1}}\right) =e^{\beta K_e}
\end{eqnarray}
or
\begin{eqnarray}
\label{corres22}
\left(e^{(\xi_e^1-\zeta_e^1)\tilde{n}_e\tau_1\tilde{n}_e^{-1}}... e^{(\xi_e^m-\zeta_e^m)\tilde{n}_e\tau_m\tilde{n}_e^{-1}}\right)= e^{\beta K_e}\,.\nonumber\\
\end{eqnarray}
Further, we have
\begin{eqnarray}
\label{corres3}
K_e =\frac{1}{\beta}n_e((\xi_e^1-\zeta_e^1)\tau_1+...+(\xi_e^m-\zeta_e^m)\tau_m)n_e^{-1}
\end{eqnarray}
or
\begin{eqnarray}
\label{corres32}
K_e =\frac{1}{\beta}\tilde{n}_e((\xi_e^1-\zeta_e^1)\tau_1+...+(\xi_e^m-\zeta_e^m)\tau_m)\tilde{n}_e^{-1}
\end{eqnarray}
when it is expressed in the perspectives of  the source point and target point of $e$ respectively.

The set of the variables $((\eta^e_1,...,\eta^e_m), (\xi^e_1,...,\xi^e_m),\mathbb{V}_e, \tilde{\mathbb{V}}_e)$ gives the generalization of twisted geometry parametrization for the $SO(D+1)$ holonomy-flux phase space. Comparing with the twisted geometry parametrization for the edge-simplicity constraint surface in the $SO(D+1)$ holonomy-flux phase space introduced in our companion paper \cite{Long:2020tlr}, this generalized parametrization scheme covers the full $SO(D+1)$ holonomy-flux phase space. We will now carry out an analysis of the symplectic structure of the $SO(D+1)$ holonomy-flux phase space based on the variables $((\eta^e_1,...,\eta^e_m), (\xi^e_1,...,\xi^e_m),\mathbb{V}_e, \tilde{\mathbb{V}}_e)$ , before coming back to provide more support on the relation between the generalized parametrization scheme in this paper and that only for the edge simplicity constraint surface given in our companion paper \cite{Long:2020tlr}.

\section{Symplectic analysis of $SO(D+1)$ holonomy-flux phase space}\label{sec4}
Notice that the discussions in this section only depend on each single edge of the graph. To simplify our notations,  we will focus on the analysis on a single edge and omit the label $e$ without loss of generality.
\subsection{Symplectic structure of $SO(D+1)$  holonomy-flux phase space}

The symplectic structure of $SO(D+1)$  holonomy-flux phase space has been discussed in our companion paper \cite{Long:2020tlr}, let us give a brief review of the main notations as follows.
Recall that the $SO(D+1)$  holonomy-flux phase space associated with each edge of a given graph can be given by the group cotangent space $T^*SO(D+1)$, as a phase space it enjoys the natural symplectic structure of the $T^*SO(D+1)$. To give the explicit formulation of this symplectic structure, let us introduce the function $f(h)$ on $SO(D+1)\ni h$, and the element $p_X\in so(D  +1)^\ast$ which is a linear function of $Y\in so(D+1)$ defined by
\begin{equation}
p_X(Y)\equiv X^{KL}Y_{KL},
\end{equation}
where $X=X^{KL}\in so(D+1)$.
A right-invariant vector field $\hat X$ associated to the Lie algebra element $X\in so(D+1)$, acts on a function $f(h)$ via the right derivative $\nabla_X^R$ as
\begin{equation}
\nabla_X^Rf(h)\equiv \frac{d}{dt}f(e^{-tX}h)|_{t=0};
\end{equation}
under the adjoint transformation $X\mapsto -hXh^{-1}$, we obtain the corresponding left derivative
\begin{equation}
\nabla_X^Lf(h)\equiv \frac{d}{dt}f(he^{tX})|_{t=0}=-\nabla^R_{hXh^{-1}}f(h).
\end{equation}
One can straightforwardly show that the map from the right invariant vector fields $\hat{X}$ to the corresponding elements $X\in so(D+1)$  is given by the algebra-valued, right-invariant 1-form $dhh^{-1}$, which reads
\begin{equation}
i_{\hat{X}}(dhh^{-1})=(\mathcal{L}_{\hat{X}}h)h^{-1}=-X,
\end{equation}
where $i$ denotes the interior product, and $\mathcal{L}_{\hat{Y}}\equiv i_{\hat{Y}}d+di_{\hat{Y}}$ denotes the Lie derivative.
Now, the natural symplectic potential for $T^\ast SO(D+1)$ can be expressed as
\begin{eqnarray}\label{sympotential}
\Theta\equiv X^{IJ}(dhh^{-1})_{IJ} \equiv\textrm{Tr}(Xdhh^{-1}).
\end{eqnarray}
The symplectic 2-form then follows as
\begin{equation}
\Omega\equiv -d\Theta=- d\textrm{Tr}(Xdhh^{-1})=\frac{1}{2}\textrm{Tr}(d\tilde{X}\wedge h^{-1}dh-dX\wedge dhh^{-1})
\end{equation}
where we have introduced $\tilde{X}\equiv-h^{-1}Xh$. From the symplectic 2-form, the Poisson brackets among the interesting phase space functions $f\equiv f(h)$ and $p_Y\equiv p_Y(X)=Y^{IJ}X_{IJ}$ is given by \cite{Long:2020tlr}
\begin{equation}\label{Poiss}
\{p_Y,p_Z\}=p_{[Y,Z]},\quad \{p_Y,f(h)\}=\nabla^R_Yf(h),\quad \{f(h),f'(h)\}=0.
\end{equation}
One can see from the brackets (\ref{Poiss}) that the Poisson action of $p_Y(X)$ generates left derivatives. Similarly, it is easy to check that the action of $\tilde{p}_Y(X)\equiv Y^{IJ}\tilde{X}_{IJ}$ with $\tilde{X}=-h^{-1}Xh$ generate the right derivative $\{\tilde{p}_Y,f(h)\}=\nabla^L_Yf(h)$. Moreover, one can check the commutative relation $\{p_Y,\tilde{p}_Z\}=0$. Finally, it is easy to verify that, by setting $2\kappa/a^{D-1}=1$, the Poisson brackets (\ref{Poiss}) given by the  natural symplectic potential \eqref{sympotential} for $T^\ast SO(D+1)$ are identical with the one \eqref{hXP1} induced by the symplectic structure \eqref{Poisson1} in the $SO(D+1)$ connection phase space \cite{Long:2020tlr}. In the following part of this article, we will analyze the symplectic structure on $T^\ast SO(D+1)$ based on the symplectic potential $\Theta$ without loss of generality.


\subsection{ Symplectomorphism between $SO(D+1)$ holonomy-flux phase space and generalized twisted geometry parameter space}

From now on, let us focus on the analysis on one single edge $e$ of given graph $\gamma$, and we omit the the label $e$ for all of the notations.
Denote by $B:=\mathbb{Q}_m\times \tilde{\mathbb{Q}}_m \times (\times_{\imath=1}^m \mathbb{R}^\imath_+)\times\mathbb{T}^m$  the collection of the generalized twisted geometric parameters $(\mathbb{V},\tilde{\mathbb{V}},\vec{\chi},\vec{\xi})$. 
It is easy to see that the map \eqref{para} is not a one to one mapping. In fact, the map \eqref{para} is a many to one mapping on the boundary of $B$ defined by $\eta^m= 0$.
One can decompose $B=B_0\cup\dot{B}$ with
\begin{equation}
\dot{B}:= B|_{|\eta_m|> 0}
\end{equation}
and
\begin{equation}
B_0:= B\setminus \dot{B}.
\end{equation}
Then, one can find that the map \eqref{para} is a one to one mapping between $\dot{B}$ and its image $\dot{B}^\ast\subset T^\ast SO(D+1)$, while it is a many to one mapping between
 $B_0$  and its image $B_0^\ast \subset T^\ast SO(D+1)$. We will first focus on the symplectic structure on $B$ in this subsection, and then go back to consider the many to one mapping between
 $B_0$  and its image $B_0^\ast $ in section \ref{4.5}.

The one to one mapping between $\dot{B}$ and its image $\dot{B}^\ast\subset T^\ast SO(D+1)$ is also an isomorphism
\begin{equation}\label{PSO}
\dot{B}\rightarrow \dot{B}^\ast\subset T^\ast SO(D+1).
\end{equation}
Based on the isomorphism \eqref{PSO}, we may use the  generalized twisted geometric parameters  to express the induced symplectic structure of $ \dot{B}^\ast\subset T^\ast SO(D+1)$ inherited from the phase space $T^\ast SO(D+1)$. First, the induced symplectic potential can be expressed as
\begin{eqnarray}
\Theta_{ \dot{B}^\ast}&=& \textrm{Tr}(Xdhh^{-1})|_{\dot{B}^\ast\subset T^\ast SO(D+1)}\\\nonumber
&=& \frac{1}{2}\sum_{\imath'=1}^m\eta_{\imath'}\textrm{Tr}(n\tau_{\imath'}n^{-1} (dnn^{-1}+n(\sum_{\imath}d\xi^\imath\tau_\imath)n^{-1}-ne^{\sum_{\imath}\xi^\imath\tau_\imath} \tilde{n}^{-1}d\tilde{n}\tilde{n}^{-1}\tilde{n}e^{-\sum_{\imath}\xi^\imath\tau_\imath} n^{-1})) \\\nonumber
&=& \frac{1}{2}\sum_{\imath=1}^m\eta_{\imath}\textrm{Tr}(V_\imath dnn^{-1})+ \sum_{\imath=1}^m\eta_{\imath}d\xi^\imath- \frac{1}{2}\sum_{\imath=1}^m\eta_{\imath}\textrm{Tr}(\tilde{V}_\imath d\tilde{n}\tilde{n}^{-1}).
\end{eqnarray}
In the space $B$, one can extend the potential $\Theta_{ \dot{B}}=\Theta_{ \dot{B}^\ast}$ in the limit $|\eta_m|\to0$ and  define
\begin{equation}
\Theta_{B}\equiv \frac{1}{2}\sum_{\imath=1}^m\eta_{\imath}\textrm{Tr}(V_\imath dnn^{-1})+ \sum_{\imath=1}^m\eta_{\imath}d\xi^\imath- \frac{1}{2}\sum_{\imath=1}^m\eta_{\imath}\textrm{Tr}(\tilde{V}_\imath d\tilde{n}\tilde{n}^{-1})
\end{equation}
as the symplectic potential on $B$. This potential gives the symplectic form $\Omega_B$ as
\begin{eqnarray}
\Omega_B=-d\Theta_B &=& \frac{1}{2}\sum_{\imath=1}^m\eta_{\imath}\textrm{Tr}(V_\imath dnn^{-1}\wedge dnn^{-1})-\frac{1}{2}\sum_{\imath=1}^m\eta_{\imath}\textrm{Tr}(\tilde{V}_\imath d\tilde{n}\tilde{n}^{-1}\wedge d\tilde{n}\tilde{n}^{-1}) \\\nonumber
&& -\sum_{\imath=1}^md\eta_\imath\wedge (d\xi_\imath+\frac{1}{2}\textrm{Tr}(V_\imath dnn^{-1})-\frac{1}{2}\textrm{Tr}(\tilde{V}_\imath d\tilde{n}\tilde{n}^{-1})).
\end{eqnarray}
It is clear that in the $\eta_m=0$ region of the above (pre-)symplectic structure is degenerate, as expected due to the degeneracy in the parametrization itself in the ${\eta_m= 0}$ region of $T^\ast SO(D+1)$.

 We are interested in the Poisson algebras between these twisted-geometry variables using the presymplectic form $\Omega_B$. In order to give the explicit Poisson brackets, in the following section we will study the Hopf sections $n(\mathbb{V})$ and $\tilde{n}(\tilde{\mathbb{V}})$ in the perspectives of their contributions to the Hamiltonian fields on $B$ defined by $\Omega_B$ .

\subsection{Geometric action on the Hopf section and its decomposition}

\subsubsection{Geometric action on the Hopf section}\label{4.2}

The Hopf map is defined as a special projection map $\pi: SO(D+1)\mapsto \mathbb{Q}_{m}$ with $\mathbb{Q}_{m}:=SO(D+1)/\mathbb{T}^m$, such that every element in $\mathbb{Q}_{m}$ comes from an orbit generated by the maximal subgroup $\mathbb{T}^m$ of  $SO(D+1)$ that fixed all of the elements in the set $\{\tau_1,\tau_2,...,\tau_m\}$. In the definition representation of $SO(D+1)$ the Hopf map reads
\begin{eqnarray}
\pi: \quad SO(D+1) &\rightarrow& \mathbb{Q}_{m} \\\nonumber
g &\rightarrow& \mathbb{V}(g)=(g\tau_1g^{-1}, g\tau_2g^{-1},...).
\end{eqnarray}
Note that $\mathbb{V}(g)$ is invariant under $g\mapsto g^{\alpha_1,\alpha_2,...,\alpha_m}=ge^{\alpha_1\tau_1+\alpha_2\tau_2+...\alpha_m\tau_m}$, thus it is a function of $\frac{D(D+1)}{2}-[\frac{D+1}{2}]$ variables only. This result shows that $SO(D+1)$ can be seen as a bundle (which is referred to as Hopf bundle) over $\mathbb{Q}_{m}$ with the $\mathbb{T}^m$
fibers. On this bundle we can introduce the Hopf sections, each as an inverse map to the above projection
\begin{eqnarray}
n:\quad \mathbb{Q}_{m} &\rightarrow& SO(D+1)\\\nonumber
\mathbb{V}&\mapsto& n(\mathbb{V}),
\end{eqnarray}
such that $\pi(n(\mathbb{V}))=\mathbb{V}$. This section assigns a specific $SO(D+1)$ element $n$ to each member of the $\mathbb{Q}_{m}$, and it is easy to see that any given section $n$ is related to all other sections via $n^{\alpha_1,\alpha_2,...,\alpha_m}\equiv ne^{\alpha_1\tau_1+\alpha_2\tau_2+...\alpha_m\tau_m}$; hence the free angles $\{\alpha_1,\alpha_2,...,\alpha_m\}$ parametrize the set of all possible Hopf sections.


Notice that each algebra element $X\in so(D+1)$ can be associated to a vector field $\hat{X}$ on $\mathbb{Q}_{m}$, which acts on a function $f(\mathbb{V})$ of $\mathbb{Q}_{m}$ as
\begin{equation}
\mathcal{L}_{\hat{X}}f(\mathbb{V}):=\frac{d}{dt}f(e^{-tX}\mathbb{V}e^{tX})|_{t=0},
\end{equation}
where $g\mathbb{V}g^{-1}:=(gV_1g^{-1}, gV_2g^{-1},...,gV_mg^{-1})$ with $g\in SO(D+1)$. Similarly, for a $so(D+1)$ valued function $S=S(\mathbb{V})$ on $\mathbb{Q}_{m}$, it can be also associated to a vector field $\hat{S}$ on $\mathbb{Q}_{m}$, , which acts on the function $f(\mathbb{V})$ of $\mathbb{Q}_{m}$ as
\begin{equation}
\mathcal{L}_{\hat{S}}f(\mathbb{V}):=\frac{d}{dt}f(e^{-tS}\mathbb{V}e^{tS})|_{t=0}.
\end{equation}
Specifically, for the linear functions we have
\begin{equation}\label{LXV}
\mathcal{L}_{\hat{X}}\mathbb{V}:=(\mathcal{L}_{\hat{X}}V_1,..., \mathcal{L}_{\hat{X}}V_m)=(-[X,V_1],...,-[X,V_m])=:-[X,\mathbb{V}].
\end{equation}
 Especially, we are interested in the action of the vector fields on the Hopf section $n$. Notice that we have
\begin{equation}
\label{LX}
\mathcal{L}_{\hat{X}}V_\imath(n)=(\mathcal{L}_{\hat{X}}n)\tau_\imath n^{-1} +n\tau_\imath(\mathcal{L}_{\hat{X}}n^{-1})=[(\mathcal{L}_{\hat{X}}n)n^{-1}, V_\imath], \quad  \forall \imath\in\{1,...,m\}.
\end{equation}
Comparing this result with (\ref{LX}), we deduce that
\begin{equation}\label{LXuu}
(\mathcal{L}_{\hat{X}}n)n^{-1}=-X+\sum_{\imath}V_\imath F^\imath_X(\mathbb{V}),
\end{equation}
where $F^\imath_X(\mathbb{V})$ are functions on $\mathbb{Q}_{m}$, so that $V_\imath F^\imath_X(\mathbb{V})$ commuting with the element $\mathbb{V}$ for all $\imath$.\\ \\
\textbf{Lemma.}
The solution functions $L_\imath^{IJ}\equiv L^\imath: \mathbb{Q}_{m}\mapsto so(D+1)$ of the equations
\begin{equation}\label{Lemma1}
\textrm{Tr}(L^\imath dnn^{-1})=0, \quad L_\imath^{IJ}V_{\imath',IJ}=\delta_{\imath,\imath'},
\end{equation}
appears in the Lie derivative of the Hopf map section $n(\mathbb{V})$ as,
\begin{equation}\label{LF}
L^\imath_X:=L^{IJ}_\imath X_{IJ}=F^\imath_X
\end{equation}
and it satisfies the key coherence identity
\begin{equation}\label{LXY}
\mathcal{L}_{\hat{X}}L^\imath_Y-\mathcal{L}_{\hat{Y}}L^\imath_X=L^\imath_{[X,Y]}.
\end{equation}
Finally, the general solution to this identity satisfying the conditions $L_\imath^{IJ}V_{\imath',IJ}=\delta_{\imath,\imath'}$ is given by
\begin{equation}\label{gf}
L'^\imath_X=L^\imath_X+\mathcal{L}_{\hat{X}}\alpha^\imath
\end{equation}
where $\alpha^\imath$ is a function on $\mathbb{Q}_{m}$.

\textbf{Proof.}

To prove Eq.\eqref{LF}, let us take the interior product of an arbitrary vector field $\hat{X}$ with the definition $\textrm{Tr}(L^\imath dnn^{-1})=0$ and consider $(\mathcal{L}_{\hat{X}}n)n^{-1}=i_{\hat{X}}(dnn^{-1})$ given by the definition of Lie derivative, we have
\begin{equation}
0=i_{\hat{X}}\textrm{Tr}(L^\imath dnn^{-1})=\textrm{Tr}(L^\imath(\mathcal{L}_{\hat{X}}n)n^{-1}) =-\textrm{Tr}(L^\imath X)+\sum_{\imath'=1}^mF^{\imath'}_X\textrm{Tr}(L^\imath V_{\imath'})=-L^\imath_X+F^\imath_X,
\end{equation}
where we used $\textrm{Tr}(L^\imath V_{\imath'})=L_\imath^{IJ}V_{\imath',IJ}=\delta_{\imath,\imath'}$ and (\ref{LXuu}). Thus, we proved $F^\imath_X=L^\imath_X$.

To prove Eq.(\ref{LXY}), we first consider that
\begin{eqnarray}
\mathcal{L}_{\hat{X}}(dnn^{-1})&=&i_{\hat{X}}(dnn^{-1}\wedge dnn^{-1})+d[(\mathcal{L}_{\hat{X}}n)n^{-1}]\\\nonumber
&=&[-X+\sum_{\imath}V_\imath L^\imath_X,dnn^{-1}]+d(-X+\sum_{\imath}V_\imath L^\imath_X)\\\nonumber
&=&\sum_{\imath}V_\imath dL^\imath_X-[X,dnn^{-1}],
\end{eqnarray}
where we used the definition of Lie derivative in the first equality, Eq.(\ref{LXuu}) in the second and $dV_\imath=[dnn^{-1},V_\imath]$ in the third. Then, the above equation leads to
\begin{equation}\label{kkklll}
0=\mathcal{L}_{\hat{X}}\textrm{Tr}(L^\imath dnn^{-1}) =\textrm{Tr}((\mathcal{L}_{\hat{X}}L^\imath-[L^\imath,X])dnn^{-1}) +dL^\imath_X
\end{equation}
by using the equalities $\textrm{Tr}(L^\imath V_{\imath'})=\delta_{\imath,\imath'}$.
Further, let us take the interior product of Eq.\eqref{kkklll} with $\hat{Y}$ and we get
\begin{eqnarray}
\mathcal{L}_{\hat{Y}}L^\imath_X&=& \textrm{Tr}((\mathcal{L}_{\hat{X}}L^\imath-[L^\imath,X] )(Y-\sum_{\imath'}V_{\imath'} L^{\imath'}_Y))\\\nonumber
&=&\mathcal{L}_{\hat{X}}L^\imath_Y-L^\imath_{[X,Y]}-\sum_{\imath'}L^{\imath'}_Y\left(\textrm{Tr}((\mathcal{L}_{\hat{X}}L^\imath)V_{\imath'}) -\textrm{Tr}(L^\imath[X,V_{\imath'}])\right)\\\nonumber
&=&\mathcal{L}_{\hat{X}}L^\imath_Y-L^\imath_{[X,Y]}-\sum_{\imath'}L^{\imath'}_Y\mathcal{L}_{\hat{X}}(\textrm{Tr}(L^\imath V_{\imath'}) ),
\end{eqnarray}
where the last term vanishes, thus we obtain the coherence identity (\ref{LXY}).

To show Eq.(\ref{gf}),  let us suppose that we have another solution $L'^\imath$ to the coherence identity and also the condition $\textrm{Tr}(L'^\imath V_{\imath'})=L'^{IJ}_\imath V_{\imath',IJ}=\delta_{\imath,\imath'}$. Considering the 1-form $\phi^\imath\equiv -\textrm{Tr}(L'^\imath dnn^{-1})$, one can see that its contraction with $\hat{X}$
\begin{equation}
\label{Ltransf}
\phi^\imath_X\equiv i_{\hat{X}}\phi^\imath=-\textrm{Tr}(L'^\imath (\mathcal{L}_{\hat{X}}n)n^{-1})=L'^\imath _X-L^\imath_X
\end{equation}
is the difference between the two solutions $L'^\imath _X$ and $L^\imath_X$. Thus, $\phi^\imath_X$ is also a solution to the coherence identity \eqref{LXY}. This result together with the
definition of the differential $i_{\hat{X}}i_{\hat{Y}}d\phi^\imath=\mathcal{L}_{\hat{Y}}\phi^\imath_X -\mathcal{L}_{\hat{X}}\phi^\imath_Y+\phi^\imath_{[X,Y]}$ implies that $d\phi^\imath=0$, which means that there exists a function $\alpha^\imath$ locally at least, such that $\phi^\imath=d\alpha^\imath$ and thus $L'^\imath_X=L^\imath_X+\mathcal{L}_{\hat{X}}\alpha^\imath$. This proves the Eq. (\ref{gf}).

$\square$

Finally, let us recall that the freedom in choosing the Hopf section lies in the function parameters $\alpha^\imath(\mathbb{V})$ in the expression $n'(\mathbb{V})\equiv n(\mathbb{V})e^{\sum_{\imath}\alpha^\imath(\mathbb{V})\tau_\imath}$ for all possible choices of the sections. By applying Eq.\eqref{LXuu} to this $n'$, we immediately get $L'^\imath_X= L^\imath_X+ i_{\hat{X}}d\alpha^\imath$. Referring to \eqref{Ltransf}, we can conclude that the function $L^\imath$ is exactly the function coefficient for the component of $(dn)n^{-1}$ in the $V_\imath$ direction, which is determined by a choice of the Hopf section $n$.

\subsubsection{Decomposition and sequence of the Hopf section }\label{hopfdecom}

As we will see in  following part of this article, the Hopf section $n$ and the geometric action on it are closely related to the symplectic structure and the symplectic reduction on $B$. To analyze the Hopf section $\mathbb{Q}_{m}$ more explicitly, let us consider the decomposition of the Hopf section $n$. Recall the definition $\mathbb{Q}_m:=SO(D+1)/\mathbb{T}^m$, one can decompose $\mathbb{Q}_{m}$ as
\begin{equation}
\mathbb{Q}_{m}=\mathbb{D}_1\times\mathbb{D}_2\times...\times\mathbb{D}_{m}
\end{equation}
with
 \begin{equation}\label{D1}
 \mathbb{D}_1:=SO(D+1)/(SO(2)_{\tau_1}\times SO(D-1)_{[\tau_1]}),
 \end{equation}
 \begin{equation}
 \mathbb{D}_2:=SO(D-1)_{[\tau_1]}/(SO(2)_{\tau_2}\times SO(D-3)_{[\tau_2]}),
 \end{equation}
 \begin{equation}
 ...
 \end{equation}
 \begin{equation}
 \mathbb{D}_{m}:=SO(D+3-2m)_{[\tau_{(m-1)}]}/SO(2)_{\tau_m},
 \end{equation}
where $SO(2)_{\tau_\imath}$ is the group generated by $\tau_\imath$ and $SO(D+1-2\imath)_{[\tau_{\imath}]}$ is the maximal subgroup of $SO(D+1)$ which preserves $(\tau_{1},...,\tau_\imath)$ and has the Cartan subalgebra spanned by $(\tau_{(\imath+1)},...,\tau_m)$. Here one should notice that both of $SO(2)_{\tau_\imath}$ and $SO(D+1-2\imath)_{[\tau_{\imath}]}$ preserve $(\tau_1,...,\tau_\imath)$. Then, the
Hopf section $n$ can be decomposed as
\begin{equation}
n=n_1n_2...n_m.
\end{equation}
 This decomposition gives a sequence of the Hopf sections, which reads
\begin{equation}\label{hopfs}
n_1,\ n_1n_2,\ n_1n_2n_3,\ ...,\ n_1...n_m.
\end{equation}
For a specific one $n_1...n_\imath$ with $ \imath\in\{1,...,m\}$, it gives
\begin{eqnarray}
n_1...n_\imath: && \mathbb{D}_1\times...\times\mathbb{D}_\imath\to SO(D+1)\\\nonumber
&& (V_1,...,V_\imath)\mapsto n_1(V_1)n_2(V_1,V_2)...n_\imath(V_1,...,V_\imath),
\end{eqnarray}
where
\begin{equation}
V_1=n_1n_2...n_\imath\tau_1 n_\imath^{-1}...n_2^{-1}n_1^{-1}=n_1\tau_1 n_1^{-1},
\end{equation}
\begin{equation}
V_2=n_1n_2...n_\imath\tau_2 n_\imath^{-1}...n_2^{-1}n_1^{-1}=n_1n_2\tau_2n_2^{-1}n_1^{-1},
\end{equation}
\begin{equation}
...,
\end{equation}
\begin{equation}
V_\imath=n_1n_2...n_\imath\tau_\imath n_\imath^{-1}...n_2^{-1}n_1^{-1}.
\end{equation}
Here one should notice that the decomposition $n=n_1...n_m$ is not unique. For instance, one can carry out the transformation
 \begin{equation}\label{tranhopf}
n_\imath\to n_\imath g, n_{\imath+1}\to g^{-1}n_{\imath+1}
 \end{equation}
 with $g\in SO(D+1)$ being arbitrary element which preserve $(\tau_1,...,\tau_\imath)$, and it is easy to verify that the transformation \eqref{tranhopf} preserves the Hopf section $n$ but changes $n_\imath$ and $n_{\imath+1}$ in the  decomposition $n=n_1...n_m$. We can also establish the geometric actions on the Hopf sections  $n_1$. Specifically, one can give
\begin{equation}\label{n1L}
  (\mathcal{L}_{\hat{X}}n_1)n_1^{-1}=-X+V_1\bar{L}^1_X (V_1)+\sum_{\mu}\bar{V}^\mu_1 \bar{L}^\mu_X(V_1)
\end{equation}
based on Eqs.\eqref{LXV}, \eqref{LX} and $V_1=n_1\tau_1n_1^{-1}$, where $\bar{V}^\mu_1=n_1\bar{\tau}^\mu n_1^{-1}$ with $\{\bar{\tau}^\mu\}$ being a basis of $so(D-1)_{\tau_1}$, $\bar{L}^1_X (V_1)=\bar{L}^1_{IJ}(V_1)X^{IJ}$ and  $\bar{L}^\mu_X(V_1)=\bar{L}^\mu_{IJ}(V_1) X^{IJ}$ are functions of $V_1\in \mathbb{D}_1$ \cite{Long:2020tlr}. It has been shown that
$\bar{L}^1_{IJ}(V_1)$ is the solution of the equations \cite{Long:2020tlr}
\begin{equation}\label{barL1}
\text{Tr}(\bar{L}^1 dn_1 n_1^{-1})=0,\quad \text{Tr}(\bar{L}^1V_1)=1,\ \text{and} \  \text{Tr}(\bar{L}^1 \bar{V}^\mu_1)=0, \forall \mu.
\end{equation}
By comparing Eq.\eqref{barL1} and  Eq.\eqref{Lemma1}, it is easy to see that $L^1=\bar{L}^1$ is a solution of $L^1$ in Eq.\eqref{Lemma1}. This result will be a key ingredient in discussions in the next section.

Now, by applying the results of this section to the presymplectic form $\Omega_B$, we will identify the Hamiltonian fields in $B$ and compute the Poisson brackets.

\subsection{Computation of Hamiltonian vector fields in pre-symplectic manifold $B$}

Let us recall the pre-symplectic potential $\Theta_{B}\equiv \frac{1}{2}\sum_{\imath=1}^m\eta_{\imath}\textrm{Tr}(V_\imath dnn^{-1})+ \sum_{\imath=1}^m\eta_{\imath}d\xi^\imath- \frac{1}{2}\sum_{\imath=1}^m\eta_{\imath}\textrm{Tr}(\tilde{V}_\imath d\tilde{n}\tilde{n}^{-1})$ induced from the $SO(D+1)$ holonomy-flux phase space, which defines the pre-sympletic form $\Omega_B$ as
\begin{eqnarray}
\Omega_B=-d\Theta_B &=& \frac{1}{2}\sum_{\imath=1}^m\eta_{\imath}\textrm{Tr}(V_\imath dnn^{-1}\wedge dnn^{-1})-\frac{1}{2}\sum_{\imath=1}^m\eta_{\imath}\textrm{Tr}(\tilde{V}_\imath d\tilde{n}\tilde{n}^{-1}\wedge d\tilde{n}\tilde{n}^{-1}) \\\nonumber
&& -\sum_{\imath=1}^md\eta_\imath\wedge \left(d\xi_\imath+\frac{1}{2}\textrm{Tr}(V_\imath dnn^{-1})-\frac{1}{2}\textrm{Tr}(\tilde{V}_\imath d\tilde{n}\tilde{n}^{-1})\right).
\end{eqnarray}

The associated Poisson brackets can be calculated by considering the Hamiltonian vector fields on $B$. Let us denote the Hamiltonian vector field for the function $f$ as $\psi_f$ , where $f\in \{\eta_\imath, \xi_\imath, p_X\equiv \frac{1}{2}\sum_{\imath}\eta_\imath V^\imath_X=\frac{1}{2}\sum_{\imath}\eta_\imath V^\imath_{IJ}X^{IJ}, \tilde{p}_X\equiv \frac{1}{2}\sum_{\imath}\eta_\imath\tilde{V}^\imath_X=\frac{1}{2}\sum_{\imath}\eta_\imath\tilde{V}^\imath_{IJ}X^{IJ}\}$. Then, using  $i_{\psi_f}\Omega_B=-df$, the vector fields could be checked to be given by
\begin{eqnarray}\label{vf}
\psi_{p_X} &=& \hat{X}-\sum_{\imath}L^\imath_X(\mathbb{V})\partial_{\xi_\imath},\quad \psi_{\tilde{p}_X} = - \hat{\tilde{X}}-\sum_{\imath}L^{\imath}_X(\tilde{\mathbb{V}})\partial_{\xi_{\imath}}, \quad \psi_{\eta_\imath}= -\partial_{\xi_\imath}. 
\end{eqnarray}
Here $\hat{X}$ are the vector fields generating the adjoint action on $\mathbb{Q}_{m}$ labelled by $\mathbb{V}$, associated to the algebra elements $X$. Similarly, $\hat{\tilde{X}}$ are the vector fields generating the adjoint action on $\mathbb{Q}_{m}$ labelled by $\tilde{\mathbb{V}}$, associated to the algebra elements $X$.\\ \\
\textbf{Proof.} The first equation of (\ref{vf}) can be checked by considering
\begin{equation}
i_{\hat{X}}\Omega_B=-\frac{1}{2}\sum_{\imath}\textrm{Tr}(d(\eta_\imath V_\imath)X)+\sum_{\imath}d\eta_\imath L^\imath_X(\mathbb{V}).
\end{equation}
Notice that we have $i_{\partial_{\xi_\imath}}\Omega_B=d\eta_\imath$, the first equation of (\ref{vf}) follows immediately. The computation for $\psi_{\tilde{p}_X}$ can be carried out similarly, with an opposite sign due to the reversal of the orientation. 
\\$\square$

\subsection{Reduction of the pre-symplectic manifold $B$}\label{4.5}

Recall that in the $\eta_m=0$ region $\Omega_B$ is degenerate, as expected due to the degeneracy of the parametrization \eqref{para} in the ${\eta_m= 0}$ region.
Let us now address this degeneracy to get a true symplectic manifold. We can reduce the pre-symplectic manifold $B$ with respect to the vector fields $\hat{E}$ in the kernel of $\Omega_B$, i.e. to consider the quotient manifold $\bar{B}\equiv B/\textrm{Ker}(\Omega_B)$. The result would be a symplectic manifold with non-degenerate 2-form given by the quotient projection of $\Omega_B$.

 In obtaining the space $\bar{B}$, we can introduce the equivalence classes under the equivalence relation $p\sim p'$ whenever $p'=e^{\hat{E}}p$, with $\hat{E}\in \textrm{Ker}(\Omega_B)$ and $p, p'\in B$. The operation is thus determined by the vector fields in the kernel of $\Omega_B$. Since it is obvious that the vector fields $\hat{E}\in \textrm{Ker}(\Omega_B)$ appear in the region with $\eta_m=0$, we look for the vector fields preserving the region while having the interior products with $\Omega_B$ proportional to $\eta_\imath$. Let us first consider the vector fields
\begin{equation}\label{EX1}
\hat{E}_X\equiv \psi_{p_X}-\psi_{\tilde{p}_Y},
\end{equation}
where $X\in so(D+1)$, $Y=-h^{-1}Xh$ with $h$ being a group element rotating $V^\imath$ to $\tilde{V}^\imath=-h^{-1}V^\imath h$. Indeed, using the fact that $V^\imath_X=\tilde{V}^\imath_Y$, the interior product of the field $\hat{D}_X$ with the symplectic 2-form is
\begin{equation}\label{EXOB}
i_{\hat{E}_X}\Omega_B=-\frac{1}{2}\sum_{\imath}d(\eta_\imath V^\imath_X-\eta_\imath\tilde{V}^\imath_Y)-\frac{1}{2}\sum_{\imath}\eta_\imath \textrm{Tr}(\tilde{V}^\imath dY) =-\frac{1}{2}\sum_{\imath}\eta_\imath\textrm{Tr}([V^\imath,X]dnn^{-1}).
\end{equation}
Now, let us analyze the degeneracy of $i_{\hat{E}_X}\Omega_B$. Denoted by $K^\imath$ the subspace of $B$ defined by $\eta_\imath=\eta_{\imath+1}=...=\eta_m=0$. Consider the $so(D+1)$ valued functions $F(V_1,...,V_{(\imath-1)})$ on $K^\imath$ which satisfies
 \begin{equation}\label{EX2}
n_{(\imath-1)}^{-1}...n_2^{-1}n_1^{-1}F(V_1,...,V_{(\imath-1)})n_1n_2...n_{(\imath-1)}\in so(D+3-2\imath)_{[\tau_{(\imath-1)}]},
\end{equation}
where $n_1n_2...n_{(\imath-1)}$ determined by $(V_1,...,V_{(\imath-1)})$ is from the sequence of the Hopf sections \eqref{hopfs}, $SO(D+1-2\imath)_{[\tau_{\imath}]}$  is the maximal subgroup of $SO(D+1)$ which preserves $(\tau_{1},...,\tau_\imath)$ and has the Cartan subalgebra spanned by $(\tau_{(\imath+1)},...,\tau_m)$.
Then, we can define the vector fields $\hat{E}^\imath_{F}$  by
 \begin{equation}
 \hat{E}^\imath_{F}:=\hat{E}_X|_{X=F(V_1,...,V_{(\imath-1)})},
 \end{equation}
and one can verify  $i_{\hat{E}^\imath_F}\Omega_B=0$ on $K^\imath$ by using Eq.\eqref{EXOB}. Thus, notice the relation $K^1\subset K^2\subset...\subset K^m$, we have
\begin{equation}
\text{Ker}(\Omega_B)\equiv\{\hat{E}^\imath_F| \imath\in \{1,...,m\}\}
\end{equation}
on $K^m$.



 Next, to find the equivalence class generated by the vector fields $\hat{E}^\imath_F $ on $K^\imath$, we note that the actions of the fields should rotate jointly the vectors $(V_\imath,..,V_m)$ and $(\tilde{V}_\imath,...,\tilde{V}_m)$, that is we have $\hat{E}^\imath_F (V_{\imath'})=-[F(V_1,...,V_{(\imath-1)}),V_{\imath'}]$, $\hat{E}^\imath_F(\tilde{V}_{\imath'})=-h^{-1}[F(V_1,...,V_{(\imath-1)}),V_{\imath'}]h$. Further, the actions preserves the group element $h$, since
\begin{equation}
\hat{E}_X(h)=-Xh-hY=0
\end{equation}
which ensures that $\hat{E}^\imath_F(h)=0$.
Therefore, given $p$ and $p'$ on  $K^\imath$, we have $ p'\sim p$ if and only if the two are related by a joint rotation in $(V_\imath,..,V_m)$ and $(\tilde{V}_\imath,...,\tilde{V}_m)$ and a $h$-preserving translations in $(\xi_1,...,\xi_m)$. It is easy to see that the parametrization \eqref{para} maps $ p$ and $p'\sim p$ to the same image in $T^\ast SO(D+1)$, as expected that the equivalence class generated by the vector fields $\hat{E}^\imath_F $ on $K^\imath$ also describes the degeneracy of the
parametrization \eqref{para}. After the quotient with respect to $\hat{E}^\imath_F$ on each $K_\imath$, we are left with a manifold $\bar{K}_\imath$ parametrized by only $(\eta_1,...,\eta_{(\imath-1)})$, $(V_1,...,V_{m})$, $(\tilde{V}_1,...,\tilde{V}_{(\imath-1)})$ and $(\xi_1,...,\xi_m)$. Recall that $B\equiv B|_{|\eta_m|>0}\cup K^m$ and $K^1\subset K^2\subset...\subset K^m$, let us define
\begin{equation}
\dot{K}^m:=K^m/\text{Ker}(\Omega_B)
\end{equation}
and then the quotient space $\bar{B}\equiv B|_{|\eta_m|>0}\cup \dot{K}^m$. Finally, we conclude that the parametrization \eqref{para} gives a one to one map between $\bar{B}$ and its image $T^\ast SO(D+1)$, and it can be extended as a symplectic-morphism with $\bar{B}$ being equipped with the symplectic structure $\Omega_B$.

\subsection{Poisson algebra among the twisted geometry parameters}

Based on the Hamiltonian vector fields given by the pre-symplectic potential $\Theta_{B}$, the Poisson brackets between the twisted geometry parameters can be given by
\begin{eqnarray}\label{OmegaQ}
&&\{\xi_\imath,\eta_\jmath\}=\delta_{\imath,\jmath},\nonumber\\
&&\{p_X, p_Y\}=p_{[X,Y]}, \quad \{\tilde{p}_X, \tilde{p}_Y\}=\tilde{p}_{[X,Y]}\nonumber\\
&&\{V^\imath,\eta_\jmath \}= \{\tilde{V}^{\imath},\eta_\jmath\}=0,
\end{eqnarray}
and
\begin{eqnarray}\label{brac2}
\{V^\imath,\tilde{V}^\jmath\}=0.
\end{eqnarray}
Moreover, one can show that the Poisson brackets given by $\Theta_{B}$ between $\xi_\imath$ and $p_X$, or the ones between $\xi_\imath$ and $\tilde{p}_X$ are non-trivial, and they are given by the function $L^\imath: \mathbb{Q}_{m}\rightarrow so(D+1)$ in the form
\begin{equation}\label{brac4}
\{\xi_\imath,p_X\}= L^{\imath}_X(\mathbb{V}), \quad \{\xi_\imath,\tilde{p}_X\}= L^{\imath}_X(\tilde{\mathbb{V}}),
\end{equation}
where $L^\imath_X\equiv \textrm{Tr}(L^\imath X)$ is the component of $L^\imath$ along the algebra element $X$.

Especially, the Eqs. \eqref{brac4} taken as the definition equations of the functions $L^\imath$, together with the Poisson brackets \eqref{OmegaQ}, already determined  $L^\imath$ to be exactly the results of the brackets $\{\xi_\imath,p_X\}$ and $ \{\xi_\imath,\tilde{p}_X\}$ given by the potential $\Theta_{B}$ corresponding to our choice of the Hopf sections. This result can be shown by the fact that, the function $L^\imath$ defined by Eqs.\eqref{brac4} is constrained by two conditions given by the above Poisson brackets \eqref{OmegaQ}, and these two conditions are exactly the definition of $L^\imath$ in \textbf{Lemma} in section \ref{4.2}. Let us then illustrate the details of this fact as follows. The first one of the two conditions comes from the equation
\begin{equation}
p_{IJ}L_\imath^{IJ}=p_{IJ}\{\xi_\imath,p^{IJ}\}=\frac{1}{2}\{\xi_\imath,p^{IJ}p_{IJ}\}= \frac{1}{4}\{\xi_\imath,\sum_{\jmath}\eta^2_\jmath\} =\frac{1}{2}\eta_\imath,
\end{equation}
with $p_{IJ}:=\frac{1}{2}\sum_{\jmath}(\eta_\jmath V^\jmath_{IJ})$,
which gives the normalization condition $L_\imath^{IJ}V^\jmath_{IJ}=\delta_\imath^\jmath$ in \textbf{Lemma} in section \ref{4.2}. The second one of the two conditions just comes from the Jacobi identity
\begin{equation}
\{\xi_\imath,\{p_{X},p_{Y}\}\}+\{p_{X},\{p_{Y},\xi_\imath \}\}+\{p_{Y},\{\xi_\imath,p_{X}\}\}=0,
\end{equation}
from which we get
\begin{equation}\label{coh1}
L^\imath_{[X,Y]}-\{p_{X},L_Y^\imath\}+\{p_{Y},L_X^\imath\}=0,
\end{equation}
By using
\begin{equation}
\{p_{X},L_Y^\imath\}=i_{\psi_{p_X}}dL_Y^\imath=\mathcal{L}_{\hat{X}}L_Y^\imath,
\end{equation}
 one can write the identity (\ref{coh1})  as an identity involving
Lie derivatives and we get
\begin{equation}
\mathcal{L}_{\hat{X}}L^\imath_Y-\mathcal{L}_{\hat{Y}}L^\imath_X=L^\imath_{[X,Y]},
\end{equation}
which is just the  coherence identity in \textbf{Lemma} in section \ref{4.2}.
 Now, it is easy to see these two conditions makes the \textbf{Lemma} in section \ref{4.2} applicable and we can verify the result given in the beginning of this paragraph.

\section{Relation with the twisted geometry parametrizations on edge simplicity constraint surface}\label{sec5}

The twisted geometry parametrization introduce in this article is constructed in the space $\times_{e\in\gamma}T^\ast SO(D+1)_e$, and we also have introduced the twisted geometry parametrization of the edge simplicity constraint surface $\times_{e\in\gamma}T^\ast_{\text{es}}SO(D+1)_e$ in our companion paper \cite{Long:2020tlr}. Thus, it is worth to discuss the relation between these two types of parametrizations.

We also focus on the twisted geometry parametrizations of the space $T^\ast SO(D+1)$ on a single edge without loss of generality. Then, by setting $\eta_2=...=\eta_m=0$ in Eq.\eqref{para}, we get
 \begin{equation}
 X=\frac{1}{2}\eta_1n\tau_1n^{-1}
 \end{equation}
 which parametrizes all of the simple fluxes satisfying $X^{[IJ}X^{KL]}=0$ in $so(D+1)$. Besides, recall the decomposition $n=n_1...n_m$ of the Hopf section $n$, we get
  \begin{eqnarray}\label{para22}
 X&=&\frac{1}{2}\eta_1n_1\tau_1n_1^{-1}\\\nonumber
 h&=&n_1e^{\xi^1\tau_1}\bar{n}\tilde{n}_1^{-1}
 \end{eqnarray}
 with $\bar{n}=n_2...n_me^{\xi^2\tau_2}...e^{\xi^m\tau_m}(\tilde{n}_2...\tilde{n}_m)^{-1}$. Recall the edge simplicity constraint surface $T_{\text{es}}^\ast SO(D+1)$ defined by
 \begin{equation}
 T_{\text{es}}^\ast SO(D+1)=\{(h,X)\in T^\ast SO(D+1)|X^{[IJ}X^{KL]}=0\},
 \end{equation}
  it is easy to see that  $T_{\text{es}}^\ast SO(D+1)\subset T^\ast SO(D+1)$  is parametrized by $(\eta_1,\xi_1, V_1, \tilde{V}_1, \bar{n})$ based on Eq.\eqref{para22}, where $V_1=n_1\tau_1n_1^{-1}$,  $\tilde{V}_1=\tilde{n}_1\tau_1\tilde{n}_1^{-1}$ with the Hopf sections ${n}_1$ and $\tilde{n}_1$ being given by the decompositions
 $n=n_1...n_m$ and $\tilde{n}=\tilde{n}_1...\tilde{n}_m$ respectively.
 Thus, by restricting the consideration on the edge simplicity constraint surface, the parametrization \eqref{para} reproduces the twisted geometry parametrization introduced in our companion paper \cite{Long:2020tlr}.

 We can further consider the symplectic reduction with respect to the edge simplicity constraint, which can be expressed as $\mathcal{S}_{IJKL}\equiv p_{[IJ}p_{KL]}=0$ with $p_{IJ}:=\frac{1}{2}\sum_{\imath}\eta_\imath V^\imath_{IJ}$ in twisted geometry parameters. Notice that the Hamiltonian vector field of edge simplicity constraint is spanned by
\begin{equation}\label{Hofe}
\psi^{\mathcal{S}}_{IJKL}=2p_{[IJ}(\hat{X}_{KL]}-\sum_{\imath}L^\imath_{KL]}\partial _{\xi_\imath}),
\end{equation}
where $\hat{X}_{KL}$ is the vector field generating the adjoint action of $X_{KL}$ on $\mathbb{Q}_m$ labelled by $\mathbb{V}$, with $X_{KL}$ is the $so(D+1)$ algebra element given by
$X_{KL}\equiv X^{IJ}_{KL}=\delta^{I}_{[K}\delta^{J}_{L]}$. It is easy to verify that the vector field \eqref{Hofe} only induces the transformation of holonomy on the edge simplicity constraint surface, which reads
\begin{equation}
\mathcal{L}_{\alpha^{IJKL}\psi^{\mathcal{S}}_{IJKL}}h= \frac{1}{2}\eta_1 \alpha^{IJKL}V^1_{[IJ}{\tau}_{KL]}h=  \frac{1}{2}\eta_1 \bar{\alpha}^{KL}n_1(\bar{\tau}_{KL}\bar{n})e^{\xi^1\tau_1}n_1^{-1},
\end{equation}
where $\alpha^{IJKL}$ is an arbitrary tensor satisfying $\alpha^{IJKL}=\alpha^{[IJKL]}$ and  $\bar{\alpha}^{KL}\bar{\tau}_{KL}\equiv \alpha^{IJKL}V^1_{[IJ}(n^{-1}_1{\tau}_{KL]}n_1)\in so(D-1)_{\tau_1}$. Thus, the component $\bar{n}$ is just the gauge component with respect to edge simplicity constraint. By reducing the edge simplicity constraint surface with respect to the gauge orbit generated by $\psi^{\mathcal{S}}_{IJKL}$, we get the simplicity reduced phase space $B_{\text{es}}$ given by
 \begin{equation}
B_{\text{es}}\equiv\mathbb{R}_+\times S^1\times \mathbb{D}_1\times \tilde{\mathbb{D}}_1 \equiv\{(\eta_1,\xi_1,V_1, \tilde{V}_1)\},
 \end{equation}
 where
 $\eta_1\in [0,+\infty)$, $\xi_1\in[-\pi,\pi)$, $V_1\in \mathbb{D}_1$, $ \tilde{V}_1\in\tilde{\mathbb{D}}_1$ with $\mathbb{D}_1$ and $\tilde{\mathbb{D}}_1$ are defined by Eq.\eqref{D1}.
Correspondingly, the reduced symplectic structure on $B_{\text{es}}$
gives the Poisson brackets
\begin{eqnarray}\label{Pbr}
\{\bar{p}_X, \bar{p}_Y\}= \bar{p}_{[X,Y]},\quad  \{\tilde{\bar{p}}_X, \tilde{\bar{p}}_Y\}=\tilde{\bar{p}}_{[X,Y]},\quad
\{\xi_1,\eta_1\}=1,
\end{eqnarray}
 where $\bar{p}_X\equiv \frac{1}{2}\eta_1V^1_X=\frac{1}{2}\eta_1 V^1_{IJ}X^{IJ}$ and $ \tilde{\bar{p}}_X\equiv \frac{1}{2}\eta_1\tilde{V}^1_X=\frac{1}{2}\eta_1\tilde{V}^1_{IJ}X^{IJ}$. Specifically, the Poisson bracket between $\xi_1$ and $(\bar{p}_X,  \tilde{\bar{p}}_X)$ are given by
\begin{equation}\label{Pbr1}
\{\xi_1, \bar{p}_X\}=L^1_X(\mathbb{V}),\quad  \{\xi_1, \tilde{\bar{p}}_X\}=L^1_X(\tilde{\mathbb{V}}).
\end{equation}
Notice these Poisson brackets is not independent of $(V_2,..V_m)$ and $(\tilde{V}_2,...,\tilde{V}_m)$, since $\xi_1$ contains the information of the choices of the Hopf section $n$ and $\tilde{n}$ which depend on $\mathbb{V}$ and $\tilde{\mathbb{V}}$. Recall the result of section \ref{hopfdecom},  by using the decomposition $n=n_1...n_m$ and $\tilde{n}=\tilde{n}_1...\tilde{n}_m$, one can choose the Hopf sections $n$ and $\tilde{n}$ to ensure that
\begin{equation}\label{Pbr2}
L^1(\mathbb{V})=\bar{L}^1(V_1),\ \text{and }\  L^1(\tilde{\mathbb{V}})=\bar{L}^1(\tilde{V}_1).
\end{equation}
Then, the symplectic structure on reduce phase space $B_{\text{es}}$ is given by the Eqs.\eqref{Pbr}, \eqref{Pbr1} and \eqref{Pbr2}, which is identical with that given in our companion paper \cite{Long:2020tlr}. Further, the gauge reduction with respect to Gaussian constraint and the treatment of vertex simplicity constraint can be carried out  following the same procedures as that in \cite{Long:2020tlr}.

\section{Conclusion and outlook}
The realization of gauge fixing in quantum gauge reduction and the Fermion coupling in all dimensional LQG require us to construct the coherent state in the full Hilbert space which involving the non-simple representations of $SO(D+1)$. Following previous experiences, it is reasonable to consider the generalized twisted geometry coherent state and thus it is necessary to establish the twisted geometry parametrization of the full $SO(D+1)$ holonomy-flux phase space.

We established the generalized twisted geometry parametrization for the full $SO(D+1)$ holonomy-flux phase space. In particular, the twisted geometry parameters are adapted to the splitting of the Ashtekar connection to capture the degrees of freedom of the intrinsic and extrinsic part of the spatial geometry respectively. Moreover, the symplectic structure on the $SO(D+1)$ holonomy-flux phase space is re-expressed based on the twisted geometry parameters.
Through studying the properties of the Hopf sections in $SO(D+1)$ Hopf fibre bundle, we obtained the Poisson algebra among the twisted geometry parameters. Especially, the relation between the twisted geometry parametrizations for
the edge simplicity constraint surface and the full holonomy-flux phase space $\times_{e\in\gamma}T^\ast SO(D+1)_e$ are discussed. We pointed out that the twisted geometry parametrizations for $\times_{e\in\gamma}T^\ast SO(D+1)_e$ is equivalent to that for the edge simplicity constraint surface by carrying out the gauge reduction with respect to the edge simplicity constraint, which ensures that the treatment of the anomalous vertex simplicity constraint proposed in our companion paper \cite{Long:2020tlr} are still valid for the more general case considered in this article.

The twisted geometry parametrizations for $\times_{e\in\gamma}T^\ast SO(D+1)_e$ provides us the tool which is necessary to construct the twisted geometry coherent state in the full Hilbert space of all dimensional LQG. More explicitly, similar to the construction of twisted geometry coherent state in the solution space of edge simplicity constraint, one could decompose the heat-kernel coherent state of $SO(D+1)$ based on the twisted geometry parametrization for  $\times_{e\in\gamma}T^\ast SO(D+1)_e$, and then select the terms dominated by the highest and lowest weight in each representation of $SO(D+1)$, to form the twisted geometry coherent state in the full Hilbert space of all dimensional LQG. This will be the subject of a follow up work \cite{Long:nonsimplecoh}.

It should be remarked that the twisted geometry parametrization of the $SO(D+1)$ holonomy-flux phase space are also valid for general $SO(D+1)$ (or $SO(N)$) Yang-Mills gauge theory, since they have the identical kinematical structure in loop quantization framework (up to the coupling constants). Though the ``geometry'' may be meaningless out of the framework of gravity theory, the twisted geometry parameters provide a new perspective to analyze the Poisson structure of the $SO(D+1)$ holonomy-flux phase space, which could help us to understand the quantum aspects of various $SO(D+1)$ Yang-Mills gauge theory.

\section*{Acknowledgments}
This work is supported by the National Natural Science Foundation of China (NSFC) with Grants No. 12047519, No. 11775082, No. 11875006 and No. 11961131013.

\bibliographystyle{unsrt}

\bibliography{ref}

\end{document}